\documentclass[conference]{IEEEtran}
\IEEEoverridecommandlockouts
\usepackage{cite}
\usepackage[utf8]{inputenc}   

\usepackage{amsmath,amssymb,amsfonts}
\usepackage{algorithmic}
\usepackage{graphicx}
\usepackage{balance}
\usepackage{textcomp}
\usepackage{hyperref}
\usepackage{amssymb} % 放在导言区，启用 \Letter 符号
\usepackage{xcolor}
\def\BibTeX{{\rm B\kern-.05em{\sc i\kern-.025em b}\kern-.08em
    T\kern-.1667em\lower.7ex\hbox{E}\kern-.125emX}}
\begin{document}

\title{Foe for Fraud: Transferable Adversarial Attacks in Credit Card Fraud Detection}

\title{Foe for Fraud: Transferable Adversarial Attacks in Credit Card Fraud Detection}

\author{
\IEEEauthorblockN{
Jan Lum Fok$^{a,b}$, Qingwen Zeng$^{a}$,Shiping Chen$^{c}$, Oscar Fawkes$^{b}$, Huaming Chen$^{a}$}
\IEEEauthorblockA{$^a$ The University of Sydney, Australia}
\IEEEauthorblockA{$^b$ Constantinople, Australia}
\IEEEauthorblockA{$^c$ CSIRO, Australia}
\IEEEauthorblockA{Jan@cxnpl.com, \{qingwen.zeng, huaming.chen\}@sydney.edu.au, shiping.chen@data61.csiro.au, OscarF@cxnpl.com}
}

\maketitle

\begin{abstract}
Credit card fraud detection (CCFD) is a critical application of Machine Learning (ML) in the financial sector, where accurately identifying fraudulent transactions is essential for mitigating financial losses. ML models have demonstrated their effectiveness in fraud detection task, in particular with the tabular dataset. While adversarial attacks have been extensively studied in computer vision and deep learning, their impacts on the ML models, particularly those trained on CCFD tabular datasets, remains largely unexplored. These latent vulnerabilities pose significant threats to the security and stability of the financial industry, especially in high-value transactions where losses could be substantial. To address this gap, in this paper, we present a holistic framework that investigate the robustness of CCFD ML model against adversarial perturbations under different circumstances. Specifically, the gradient-based attack methods are incorporated into the tabular credit card transaction data in both black- and white-box adversarial attacks settings. Our findings confirm that tabular data is also susceptible to subtle perturbations, highlighting the need for heightened awareness among financial technology practitioners regarding ML model security and trustworthiness. Furthermore, the experiments by transferring adversarial samples from gradient-based attack method to non-gradient-based models also verify our findings. Our results demonstrate that such attacks remain effective, emphasizing the necessity of developing robust defenses for CCFD algorithms.

\end{abstract}


\section{Introduction}

We are witnessing the widespread adoption of machine learning (ML) in financial services, which have revolutionized the areas of fraud detection, risk assessment, and anti-money laundering (AML). To date, these financial services have heavily relied on traditional ML models such as decision trees, logistic regression, and ensemble learning techniques. As a critical application of ML in financial security services, credit card fraud detection (CCFD) is an important use case which enables real-time identification of fraudulent transactions and enhancing fraud prevention strategies. Since financial institutions are now heavily relying on these ML-driven risk assessment, anomaly detection, and fraud detection systems, ensuring their robustness against adversarial threats has become a pressing concern~\cite{ref1}.

On the other hand, existing studies have demonstrated vulnerabilities in ML models trained on image data, in particular by introducing imperceptible perturbations to input data to deceive ML models. These attacks will generally leverage the structured nature of visual features to craft adversarial examples~\cite{cvref}. Many of the works focus on computer vision concerning these adversarial attacks, however it remains underexplored in finance service application such as CCFD which generally utilize tabular datasets. For tabular data, which is commonly used in financial applications, presents unique challenges due to its discrete, heterogeneous feature distributions and domain-specific constraints~\cite{ref2}. This lack of research poses a significant security risk, as financial institutions widely deploy ML-based fraud detection models without fully understanding their vulnerabilities to adversarial manipulation. Statistical reports highlight that credit card fraud caused an estimated loss of £574.2 million in the United Kingdom in 2020 alone \cite{statistics}.The extent to which established adversarial attack methods can be applied to ML models trained on tabular data for financial decision-making remains unclear, leaving a critical gap in securing ML-based fraud detection systems.% to tabular data models remains unclear.

To address this gap, this paper first examines the security of the most commonly used gradient-based ML models in the CCFD domain. We employ gradient-based attack algorithms to generate adversarial samples and investigate their impacts on fraud detection performance. To highlight broader security threats beyond traditional gradient-based architectures and uncover additional attack vectors, we investigate whether the adversarial samples can also mislead a non-gradient-based ML model. This investigation reveals critical vulnerabilities in CCFD models, demonstrating that ML models trained on tabular data are susceptible to both black-box and transfer adversarial attacks. Our findings highlight the urgent need for robust defense mechanisms to enhance the security of ML-based fraud detection systems.%critical security risks in this domain.

In summary, this paper makes following key contributions:
\begin{IEEEitemize}

    \item We present a holistic framework for transferable adversarial attacks, in particular for CCFD. Our objective is to investigate the underlying issue whereby subtle perturbations to input samples can mislead ML models in the context of fraud detection. To this end, we evaluate a gradient-based model, revealing that tabular data may serve as a susceptible attack surface for adversaries. (Sec.\ref{sec:methodology})
    \item We perform a comprehensive evaluation of the proposed framework using real-world credit card transaction datasets. Furthermore, we evaluate the effectiveness of the adversarial samples on unseen ML models, which shares entirely different model architecture and training process. In this paper, we consider a non-gradient-based ML model to assess the transferability of the adversarial attacks. With the collected results based on the attack success rate, our work demonstrates that adversarial samples generated in a white-box setting can successfully transfer to a structurally different model while maintaining high attack efficacy. (Sec.\ref{sec:experiments})

    \item Through the framework, we provide empirical insights about the security vulnerabilities of CCFD models, revealing their susceptibility to adversarial attacks, including black-box and transfer attacks. Our findings highlight the urgent need to strengthen the security of financial fraud detection systems and develop more robust defenses against adversarial threats. (Sec.\ref{sec:results_analysis})
\end{IEEEitemize}

The remainder of this paper is structured as follows: Section \ref{sec:literature_review} reviews related work. Section \ref{sec:methodology} introduces the techniques used in our proposed framework. Section \ref{sec:experiments} details the experimental setup. Section \ref{sec:results_analysis} presents the experimental results and provides an in-depth analysis of their implications. Finally, Section \ref{sec:conclusion} concludes the paper and outlines potential future research directions.

\section{Literature Review} \label{sec:literature_review}

Credit card fraud detection (CCFD) is a prominent application of machine learning (ML) in financial security, where traditional rule-based systems have been increasingly replaced by supervised models such as logistic regression, Naive Bayes, KNN, random forests, and SVM for improved detection performance~\cite{ali2022financial,34,35}. Deep neural networks have also been adopted to further enhance detection accuracy~\cite{36}. However, these models mainly focus on classification performance, while their robustness against adversarial threats remains underexplored~\cite{AMLsurvey}. Adversarial attacks---crafted perturbations designed to mislead model predictions---have been extensively studied in image-based domains using methods like the Fast Gradient Sign Method (FGSM), Projected Gradient Descent (PGD), and Carlini \& Wagner attacks~\cite{6,40,41}. However their adaptation to tabular data, especially in CCFD, poses unique challenges due to the discrete and heterogeneous nature of features~\cite{ref2}. Although some works have examined adversarial risks in CCFD, they largely concentrate on deep learning under white-box assumptions~\cite{43,44}, and penetration testing appears insufficient against such threats~\cite{lunghi2024assessing}. Meanwhile, commonly deployed traditional models like logistic regression and random forest---still favored in fraud detection for their simplicity and efficiency---have received little attention in this context~\cite{tiwari2021credit}. Moreover, the feasibility of applying transferable adversarial attacks in black-box settings for financial services remains unvalidated, forming the core research gap this work aims to address.

\section{PROPOSED FRAMEWORK} \label{sec:methodology}

The objective of this work is to investigate the security of machine learning (ML) models in credit card fraud detection (CCFD) within the financial sector. Given that this field remains relatively nascent, existing research on its security vulnerabilities is still limited. CCFD based on ML algorithms is typically formulated as a binary classification problem, where transactions are classified as either legitimate or fraudulent (e.g., credit card skimming) based on extracted features such as transaction time, transaction amount, frequency, and type. ML models will make these classifications using a predefined decision threshold. For adversarial threats, if an attacker obtains knowledge of a model's decision boundary or the inherent architecture and parameters, they can exploit this information by introducing small, strategically crafted perturbations along the gradient direction. These adversarial perturbations can alter fraud probabilities just enough to shift a fraudulent transaction below the classification threshold, causing a misclassification as a legitimate transaction. 

\begin{figure*}[htbp]
    \centering
    \includegraphics[width=.8\textwidth]{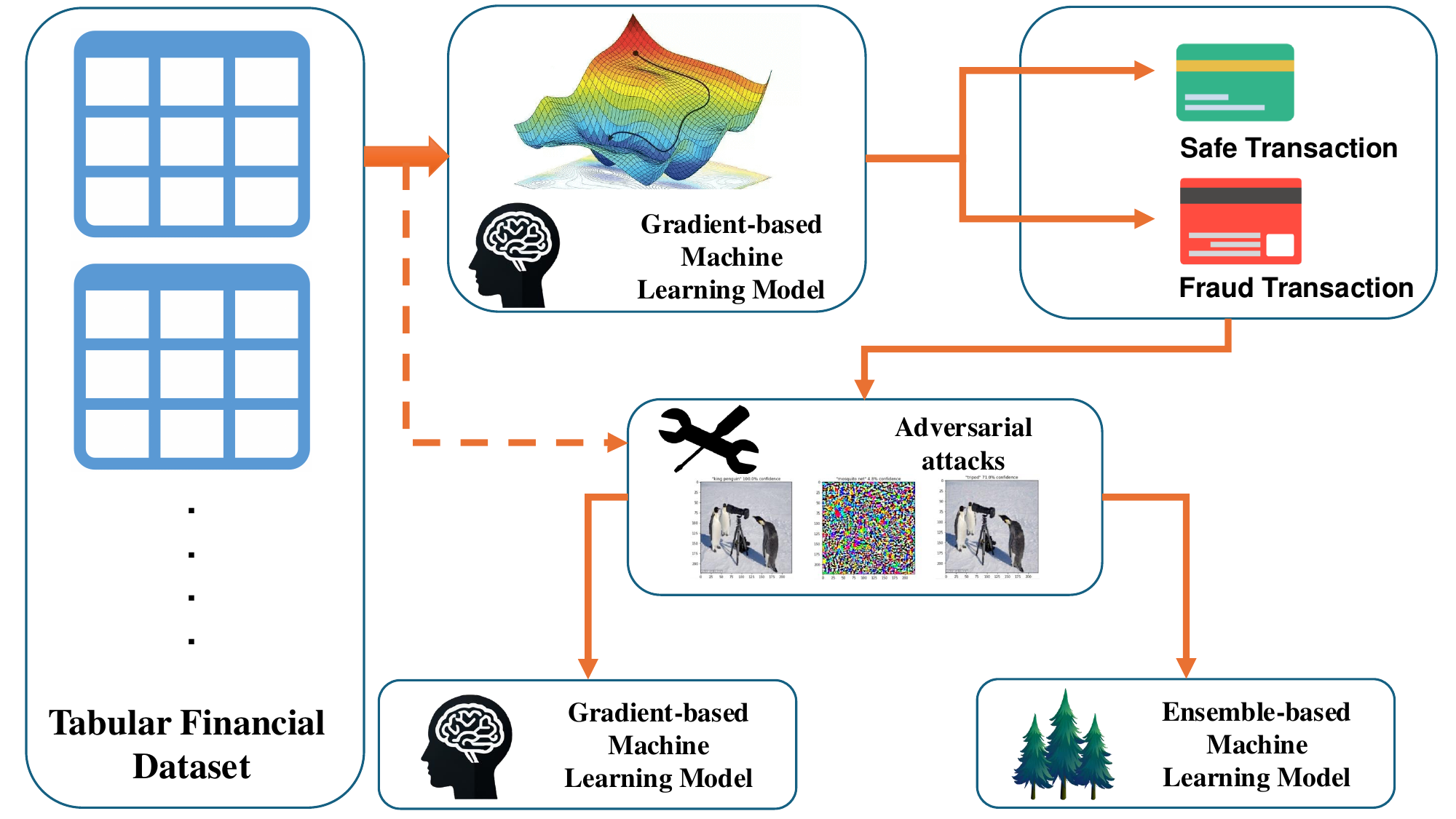}
    \caption{The overall framework of FGSM-based adversarial attack on credit card fraud detection models}
    \label{fig:framework}
\end{figure*}

In this section, we will first present the overall framework to generate the adversarial samples following by the design of white-box and black-box adversarial attacks. As shown in Fig.~\ref{fig:framework}, we first introduce an adversarial attack method, which is the Fast Gradient Sign Method (FGSM), to generate the adversarial samples for a gradient-based ML model. The goal is to evaluate whether adversarial attacks remain effective in the context of CCFD, particularly for tabular dataset. To further investigate potential attack vectors in this domain, we extract adversarial samples that successfully bypass detection in the first stage and utilize them as attack inputs against a non-gradient-based ML model. This step examines whether transferable attacks can compromise CCFD systems even when the target model's internal structure and parameters are entirely unknown.

Through this framework, we aim to uncover vulnerabilities in both ML algorithms and the underlying tabular data used in fraud detection. Our findings emphasize the necessity of integrating security measures into the development and deployment of ML-based fraud detection models in the financial sector to mitigate adversarial threats.

%In the framework as shown in Fig.~\ref{fig:framework}, we first utilize the Fast Gradient Sign Method (FGSM) to attack a gradient-based machine learning model, evaluating whether adversarial attacks remain effective in the context of credit card fraud detection, particularly in a tabular data environment. Subsequently, to further explore potential attack vectors in this field, we extract the adversarial samples that successfully bypassed detection in the first stage and use them as attack samples against a non-gradient-based machine learning model. This step examines whether transfer attacks, as a form of black-box attack, can pose a security threat to credit card fraud detection systems, even in scenarios where the internal structure and parameters of the target model are entirely unknown.

%Through this study, we aim to reveal the vulnerabilities of both machine learning algorithms and the underlying data in this domain, thereby highlighting the importance of incorporating security considerations into the development and deployment of fraud detection models in the financial sector.

\subsection{Machine learning models in our Framework}\label{sec:met_ml}
\subsubsection{Gradient-based ML}
Most of the gradient-based ML models rely on gradient descent and its variants to optimize the model parameters. For CCFD, logistic regression (LR) is a widely used gradient-based ML algorithm that estimates the probability of an event occurring based on input features~\cite{LRccfd1,LRccfd2,LRccfd3}. Unlike linear regression, which models a continuous output, LR applies the logistic function to map predictions to probabilities within the range of 0 and 1 \cite{23}. The logistic function is defined as:
\begin{equation}
    \text{Logit}(\pi) = \frac{1}{1 + \exp(-\pi)}
\end{equation}
where \(\pi\) represents the linear combination of input features and model parameters.

The model parameters are optimized using maximum likelihood estimation (MLE) to minimize classification error. Once trained, LR predicts probabilities, and classification is performed by assigning labels based on a threshold (e.g., \(\pi \geq 0.5\) is classified as class 1, otherwise class 0) \cite{23}. The sigmoid function, which characterizes LR, ensures that predictions are bounded within the probability range.

\subsubsection{Non-gradient-based ML}
A different type of ML model will be non-gradient-based models, which have entirely different architectures and training processes when compared to their gradient-based counterparts. These models generally select features and categorizes based on criteria like Gini impurity or entropy information. An example model is random forest (RF), which is an ensemble learning algorithm designed for classification and regression tasks. It constructs multiple decision trees, each trained on randomly selected subsets of the dataset and feature space \cite{26}. The randomness in RF is introduced through a process known as bootstrap aggregation (bagging), where different samples of the training data are used to build individual trees.

For classification tasks, RF determines the final prediction by applying a majority voting scheme across all decision trees. In regression tasks, it computes the average prediction from all trees. The introduction of randomness at both the data sampling and feature selection levels enhances model generalization and mitigates the risk of overfitting \cite{27}.

\subsection{Adversarial attack techniques in our framework}
While numerous adversarial attack techniques exist in the literature, this work focuses on transferable adversarial attacks in the context of CCFD. To achieve this, we employ the FGSM as the primary attack technique. FGSM, introduced by Goodfellow et al.~\cite{6}, demonstrates the vulnerability of ML models to adversarial perturbations. It generates adversarial examples by computing the gradient of the loss function with respect to the input data and applying a small perturbation in the direction of the gradient:
\begin{equation}
    \eta = \epsilon \cdot \text{sign}(\nabla_x J(\theta, x, y))
\end{equation}
where \(\epsilon\) is a small perturbation factor, \(\theta\) represents the model parameters, \(x\) is the input, \(y\) is the target label, and \(J(\theta, x, y)\) is the loss function.

The perturbed adversarial example is computed as:
\begin{equation}
    x^* = x + \eta = x + \epsilon \cdot \text{sign}(\nabla_x J(\theta, x, y))
\end{equation}
where \(x^*\) represents the generated adversarial sample. By leveraging the model’s gradient, FGSM efficiently alters input features to deceive the classifier while maintaining minimal perceptual changes.

\section{Experiments} \label{sec:experiments}
\subsection{Dataset}

We use a publicly available credit card transaction dataset from Kaggle~\cite{45}, containing 284{,}807 transactions from European cardholders over two days in 2013. The dataset is highly imbalanced, with only 0.17\% (492) labeled as fraudulent, reflecting real-world conditions. To preserve confidentiality, features were anonymized using PCA, resulting in variables V1–V28, while `Time` and `Amount` were retained. To address class imbalance, we apply the SMOTE technique~\cite{46} to oversample fraudulent cases. The dataset is split into 80\% training and 20\% testing using stratified sampling to maintain class distribution.

\subsection{Evaluation Metrics}

To comprehensively evaluate model performance on the imbalanced credit card fraud detection (CCFD) task, we report standard metrics including \textbf{Accuracy}, \textbf{Precision}, and \textbf{Recall}. Accuracy reflects overall prediction correctness, while Precision and Recall capture the model’s ability to correctly identify fraud cases and minimize false negatives. In addition, we introduce the Transferability Rate to assess how well adversarial samples generated from one model can mislead another: 
\[
\text{Transferability Rate} = \frac{\text{Misclassified Samples in Target Model}}{\text{Total Adversarial Samples}}
\]
This metric is crucial for evaluating security risks in multi-classifier fraud detection scenarios.

\subsection{Experimental Design}

We adopt logistic regression (LR) as the baseline model due to its high interpretability and relevance in financial fraud detection. Its reliance on gradient-based optimization makes it vulnerable to adversarial attacks such as the Fast Gradient Sign Method (FGSM). The model is implemented using the scikit-learn library~\cite{scikit-learn} and serves as the primary attack target.

To evaluate model robustness, FGSM attacks are applied using the Adversarial Robustness Toolbox (ART)~\cite{art} in a white-box setting with full access to model gradients. Adversarial perturbations target correctly classified fraudulent transactions in the test set, aiming to flip them to non-fraudulent labels through minimal input changes.

To assess transferability, we test whether adversarial examples crafted for LR can mislead a non-gradient-based model—random forest (RF)—trained on the same dataset with an 80/20 stratified split. This transfer-based black-box attack reveals cross-model vulnerabilities and highlights the broader risks of adversarial examples across heterogeneous fraud detection systems.

\section{Results and Analysis} \label{sec:results_analysis}
We present Fast Gradient Sign Method (FGSM) attack results on a Logistic Regression (LR) model for credit card fraud detection (CCFD), analyzing recall degradation under varying $\epsilon$ and the transferability of adversarial samples to a non-gradient-based model (Random Forest).

\subsection{Baseline Model Performance}

To establish a reference point for evaluating the impact of adversarial attacks, we first trained and evaluated a LR model on the Kaggle CCFD dataset. The evaluation metrics for the baseline model are as follows in Table~\ref{result1} :

\begin{table}[ht]
\centering
\caption{Model Evaluation Metrics}
\begin{tabular}{|c|c|}
\hline
\textbf{Metric} & \textbf{Value} \\
\hline
Accuracy & 0.99 \\
Precision & 0.17 \\
Recall & 0.92 \\
\hline
\end{tabular}
\label{result1}
\end{table}

The high recall value indicates that the model effectively detects fraudulent transactions under normal conditions. However, the relatively low precision suggests that the model produces a significant number of false positives, a trade-off that is acceptable given the priority of recall in fraud detection.

\subsection{Impact of FGSM on Logistic Regression}

To assess the vulnerability of the LR model to adversarial attacks, we generated adversarial samples using FGSM, targeting fraudulent transactions that were correctly classified in the test set. The $\epsilon$ value was set to 2.2, and the generated adversarial samples were used in place of their benign counterparts. The model’s performance on the adversarial test set is shown below in Table~\ref{result2}:

\begin{table}[ht]
\centering
\caption{Model Evaluation Metrics}
\begin{tabular}{|c|c|}
\hline
\textbf{Metric} & \textbf{Value} \\
\hline
Accuracy & 0.99 \\
Precision & 0.11 \\
Recall & 0.56 \\
\hline
\end{tabular}
\label{result2}
\end{table}

Compared to the original recall of 0.92, the recall dropped significantly, indicating that nearly 40\% of fraudulent transactions were misclassified as non-fraudulent. This significant reduction in fraud detection capability underscores the security risk posed by adversarial attacks, highlighting the need for robust defense mechanisms in CCFD systems.

\subsection{Effect of Epsilon on Model Robustness}

\begin{figure}[htbp]
    \centering
    \includegraphics[width=0.48\textwidth]{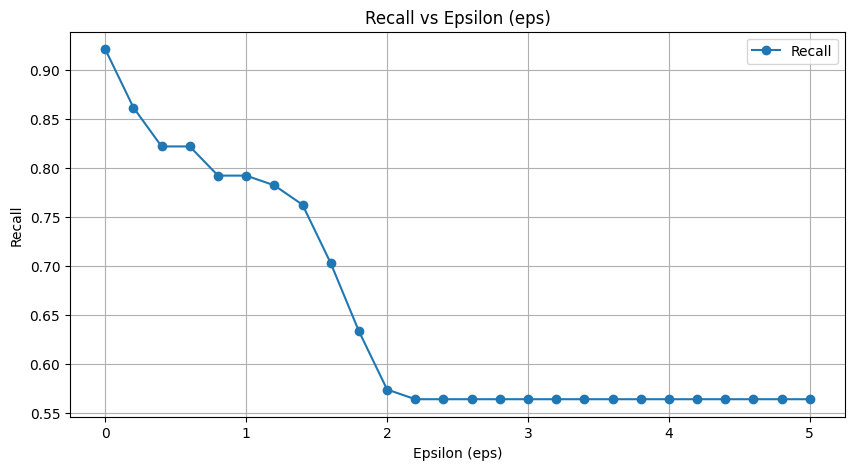}
    \caption{Effect of increasing adversarial perturbations ($\epsilon$) on model recall, showing degradation trends and eventual stabilization.}
    \label{fig:epsilon_recall}
\end{figure}

To analyze how different adversarial perturbation magnitudes affect model performance, we varied the $\epsilon$ value and observed its impact on recall. The results are plotted in Fig.~\ref{fig:epsilon_recall}.

From Fig.~\ref{fig:epsilon_recall}, we observe the following trends:
\begin{itemize}
    \item At $\epsilon = 0$, the recall is 0.92, indicating correct detection of 92\% of fraudulent transactions without adversarial perturbation.
    \item As $\epsilon$ increases, recall steadily decreases, showing the increasing effectiveness of the FGSM attack.
    \item When $\epsilon = 2.2$, recall drops to 0.56, meaning that nearly half of the fraudulent transactions are misclassified.
    \item For $\epsilon > 2.2$, recall stabilizes at 0.56, suggesting that increasing perturbation magnitude beyond this point does not further degrade model performance.
\end{itemize}

This saturation effect implies that FGSM has a maximum attack effectiveness threshold, beyond which further perturbations do not introduce additional vulnerability. One possible explanation is that as perturbations increase, they may push fraudulent samples too far from their original distribution, making them unrealistic and potentially easier to detect through anomaly detection mechanisms. Additionally, excessive perturbations could move samples beyond the model’s decision boundary in a way that no longer affects classification, leading to diminishing returns in attack effectiveness.

\subsection{Analysis of Feature Perturbations}

To better understand the effect of adversarial perturbations on individual features, we analyzed the original and adversarial values of a selected successfully perturbed fraudulent transaction. Table~\ref{table:successful_perturb_table} presents the perturbation details.

\begin{table}[ht]
\centering
\caption{Adversarial perturbation analysis for a successfully misclassified fraudulent sample.}
\begin{tabular}{|c|c|c|c|}
\hline
\textbf{Feature} & \textbf{Before (Original)} & \textbf{After (Adversarial)} & \textbf{Perturbation} \\
\hline
Time & 48884.0 & 48886.2 & 2.2 \\
V1 & -2.13905056 & -4.3390503 & -2.19999974 \\
V2 & 1.39436766 & -0.8056323 & -2.19999996 \\
V3 & -0.612034895 & -2.8120348 & -2.19999991 \\
V4 & 1.04932706 & -1.1506729 & -2.19999996 \\
V5 & -1.16210191 & -3.3621018 & -2.19999989 \\
V6 & -0.768219363 & 1.4317807 & 2.20000006 \\
V7 & -1.99723740 & 0.20276256 & 2.19999996 \\
V8 & 0.574996543 & 2.7749965 & 2.19999996 \\
V9 & -0.980831776 & 1.2191682 & 2.19999998 \\
V10 & -2.49561925 & -0.2956193 & 2.19999995 \\
V11 & 2.55558915 & 0.3555892 & -2.19999995 \\
V12 & -3.53043627 & -1.3304362 & 2.20000007 \\
V13 & -1.01623382 & 1.1837661 & 2.19999992 \\
V14 & -3.45519658 & -1.2551966 & 2.19999998 \\
V15 & -0.056363864 & 2.1436367 & 2.20000009 \\
V16 & -2.46773703 & -0.26773703 & 2.2 \\
V17 & -7.14032597 & -4.940326 & 2.19999997 \\
V18 & -1.27128002 & 0.92871994 & 2.19999996 \\
V19 & -0.00172192354 & -2.201722 & -2.20000008 \\
V20 & 0.0254265126 & 2.2254264 & 2.19999989 \\
V21 & 0.696954881 & -1.5030451 & -2.19999998 \\
V22 & 0.740003045 & -1.4599969 & -2.19999995 \\
V23 & -0.155115249 & -2.3551152 & -2.19999995 \\
V24 & -0.0506074461 & -2.2506075 & -2.20000005 \\
V25 & 0.268368293 & -1.9316317 & -2.19999999 \\
V26 & -0.469432841 & 1.7305672 & 2.20000004 \\
V27 & -0.405813768 & -2.6058137 & -2.19999993 \\
V28 & -0.152170847 & -2.352171 & -2.20000015 \\
Amount & 19.73 & 17.53 & -2.2 \\
\hline
\end{tabular}
\label{table:successful_perturb_table}
\end{table}

Key observations from Table~\ref{table:successful_perturb_table} reveal the behavior of the model under adversarial manipulation:

\begin{itemize}
    \item The PCA-transformed features (V1–V28) exhibit consistent perturbations of approximately \(\pm 2.2\), which are sufficient to mislead the logistic regression model. Since these features are derived from principal component analysis, even small changes can significantly alter the encoded representations.
    \item Named features such as \texttt{Time} and \texttt{Amount} remain nearly unchanged, indicating that adversarial perturbations target abstract feature spaces rather than raw transaction details.
    \item These findings highlight the model’s sensitivity to transformed inputs and underscore the importance of considering adversarial robustness during the preprocessing and feature engineering stages in fraud detection pipelines.
\end{itemize}

\subsection{Transferability of Adversarial Samples}

To further evaluate the impact of adversarial samples, we test their ability to transfer to another classification model. Specifically, we assess whether the adversarial samples generated using the FGSM attack on the logistic regression model can also deceive a non-gradient-based model such as Random Forest (RF). 

\subsubsection{Baseline RF Model Performance}

An RF classifier was trained on the same dataset as the LR model, following the same 80/20 train-test split. The baseline performance of RF on benign test samples is reported below in Table~\ref{result3}:

\begin{table}[ht]
\centering

\caption{Model Evaluation Metrics}
\begin{tabular}{|c|c|}
\hline
\textbf{Metric} & \textbf{Value} \\
\hline
Accuracy & 1.00 \\
Precision & 1.00 \\
Recall & 0.95 \\
\hline
\end{tabular}
\label{result3}
\end{table}

\begin{table}[ht]
\centering
\caption{Confusion matrix for baseline Random Forest model before adversarial attack.}
\begin{tabular}{|c|c|c|}
\hline
Actual & Predicted: Non-fraud & Predicted: Fraud \\
\hline
Non-fraud & 56861 & 5 \\
Fraud & 0 & 96 \\
\hline
\end{tabular}
\label{table:rf_baseline_conf_matrix}
\end{table}

As shown in Table~\ref{table:rf_baseline_conf_matrix}, RF achieves near-perfect performance in fraud detection under normal conditions.

\subsubsection{Adversarial Sample Transferability Experiment}

To examine the robustness of RF against adversarial samples, we applied the adversarial test set generated using FGSM on the LR model to the trained RF model. The results are as follows in Table~\ref{result4}:

\begin{table}[ht]
\centering
\caption{Adversarial Attack Results}
\begin{tabular}{|c|c|}
\hline
\textbf{Metric} & \textbf{Value} \\
\hline
Successful attacks & 34 \\
Failed attacks & 2 \\
Transferability success rate & 94\% \\
\hline
\end{tabular}
\label{result4}
\end{table}

The high success rate (94\%) indicates that adversarial perturbations crafted for a gradient-based model like LR can still significantly degrade the performance of a non-gradient-based model such as RF. This demonstrates that adversarial attacks are not limited to models that rely on gradients, but can also transfer across different types of classifiers.

\subsubsection{Analysis of Transferability}

Several factors contribute to the high transferability of adversarial samples:

\begin{itemize}
    \item Feature-space perturbations: Even though RF does not use gradients, it relies on feature importance for classification. Perturbations introduced by FGSM may shift the decision boundary for critical features, leading to misclassification.
    
    \item Shared decision boundaries: Both LR and RF models were trained on the same dataset, meaning they may learn similar decision boundaries. This increases the likelihood that adversarial perturbations effective for LR also impact RF.
    
    \item Structural limitations of ensemble models: While RF benefits from bagging and feature randomness, it remains susceptible to adversarial perturbations that systematically alter the input distribution, leading to incorrect classifications.
\end{itemize}

These results highlight the broader security risks of adversarial attacks in real-world fraud detection systems. Even when an organization deploys multiple classifiers to improve robustness, transferability can still expose all models to adversarial threats. Future research should explore adversarial training techniques and other defense mechanisms to mitigate these risks.

\section{Conclusion} \label{sec:conclusion}
In this paper, we have examined the vulnerability of credit card fraud detection (CCFD) models to adversarial attacks, demonstrating that adversarial samples generated via Fast Gradient Sign Method (FGSM) significantly degrade model performance. In particular, logistic regression model's recall dropped from 92\% to 56\%, and these perturbations exhibited a 94\% transferability rate to a non-gradient-based Random Forest model, underscoring the broader security risks posed by adversarial attacks in financial applications.

While various adversarial defense mechanisms have been proposed—such as adversarial training, feature regularization, and ensemble defenses~\cite{6}—most of these methods have been developed for non-tabular datasets. Gradient obfuscation methods aim to prevent attackers from exploiting model gradients~\cite{49}, while input preprocessing techniques such as noise filtering and data compression attempt to mitigate the impact of adversarial perturbations~\cite{50}.Their applicability to CCFD models remains largely unexplored. Future work should focus on adapting adversarial defenses for machine learning (ML) models on tabular datasets, evaluating their effectiveness across diverse financial datasets, and exploring hybrid defense strategies that integrate multiple mitigation techniques.

In conclusion, adversarial attacks present a systemic security risk to ML-based fraud detection, extending beyond gradient-based models. Developing robust, domain-specific adversarial defenses is crucial to ensuring the trustworthiness and resilience of financial fraud detection systems.

\section*{Acknowledgement}
The first author conducted this work at Constantinople Operating Company Pty Ltd (Constantinople), as part of a funded research collaboration with The University of Sydney. The authors would like to thank Constantinople for their support of this work.

\balance
\bibliographystyle{IEEEtran}  
\bibliography{mainContent} 

\begin{thebibliography}{10}
\providecommand{\url}[1]{#1}
\csname url@samestyle\endcsname
\providecommand{\newblock}{\relax}
\providecommand{\bibinfo}[2]{#2}
\providecommand{\BIBentrySTDinterwordspacing}{\spaceskip=0pt\relax}
\providecommand{\BIBentryALTinterwordstretchfactor}{4}
\providecommand{\BIBentryALTinterwordspacing}{\spaceskip=\fontdimen2\font plus
\BIBentryALTinterwordstretchfactor\fontdimen3\font minus \fontdimen4\font\relax}
\providecommand{\BIBforeignlanguage}[2]{{%
\expandafter\ifx\csname l@#1\endcsname\relax
\typeout{** WARNING: IEEEtran.bst: No hyphenation pattern has been}%
\typeout{** loaded for the language `#1'. Using the pattern for}%
\typeout{** the default language instead.}%
\else
\language=\csname l@#1\endcsname
\fi
#2}}
\providecommand{\BIBdecl}{\relax}
\BIBdecl

\bibitem{ref1}
\BIBentryALTinterwordspacing
D.~Lunghi, A.~Simitsis, O.~Caelen, and G.~Bontempi, ``Adversarial learning in real-world fraud detection: Challenges and perspectives,'' \emph{arXiv}, vol. 2307.01390, 2023. [Online]. Available: \url{https://doi.org/10.48550/arXiv.2307.01390}
\BIBentrySTDinterwordspacing

\bibitem{cvref}
N.~Akhtar, A.~Mian, N.~Kardan, and M.~Shah, ``Advances in adversarial attacks and defenses in computer vision: A survey,'' \emph{IEEE Access}, vol.~9, pp. 155\,161--155\,196, 2021.

\bibitem{ref2}
F.~Cartella, O.~Anunciacao, Y.~Funabiki, D.~Yamaguchi, T.~Akishita, and O.~Elshocht, ``Adversarial attacks for tabular data: Application to fraud detection and imbalanced data,'' \emph{arXiv preprint arXiv:2101.08030}, 2021.

\bibitem{statistics}
S.~Xuan, G.~Liu, Z.~Li, L.~Zheng, S.~Wang, and C.~Jiang, ``Random forest for credit card fraud detection,'' in \emph{2018 IEEE 15th International Conference on Networking, Sensing and Control (ICNSC)}, 2018, pp. 1--6.

\bibitem{ali2022financial}
A.~Ali, S.~Abd~Razak, S.~H. Othman, T.~A.~E. Eisa, A.~Al-Dhaqm, M.~Nasser, T.~Elhassan, H.~Elshafie, and A.~Saif, ``Financial fraud detection based on machine learning: a systematic literature review,'' \emph{Applied Sciences}, vol.~12, no.~19, p. 9637, 2022.

\bibitem{34}
D.~Varmedja, M.~Karanovic, S.~Sladojevic, M.~Arsenovic, and A.~Anderla, ``Credit card fraud detection - machine learning methods,'' in \emph{2019 18th International Symposium INFOTEH-JAHORINA (INFOTEH)}, 2019, pp. 1--5.

\bibitem{35}
S.~Kumar, V.~K. Gunjan, M.~D. Ansari, and R.~Pathak, ``Credit card fraud detection using support vector machine,'' in \emph{Proceedings of the 2nd International Conference on Recent Trends in Machine Learning, IoT, Smart Cities and Applications: ICMISC 2021}.\hskip 1em plus 0.5em minus 0.4em\relax Springer, 2022, pp. 27--37.

\bibitem{36}
R.~Asha and S.~K. KR, ``Credit card fraud detection using artificial neural network,'' \emph{Global Transitions Proceedings}, vol.~2, no.~1, pp. 35--41, 2021.

\bibitem{AMLsurvey}
F.~V. Jedrzejewski, L.~Thode, J.~Fischbach, T.~Gorschek, D.~Mendez, and N.~Lavesson, ``Adversarial machine learning in industry: A systematic literature review,'' \emph{Computers \& Security}, p. 103988, 2024.

\bibitem{6}
\BIBentryALTinterwordspacing
I.~J. Goodfellow, J.~Shlens, and C.~Szegedy, ``Explaining and harnessing adversarial examples,'' 2015. [Online]. Available: \url{https://arxiv.org/abs/1412.6572}
\BIBentrySTDinterwordspacing

\bibitem{40}
\BIBentryALTinterwordspacing
A.~Madry, A.~Makelov, L.~Schmidt, D.~Tsipras, and A.~Vladu, ``Towards deep learning models resistant to adversarial attacks,'' 2019. [Online]. Available: \url{https://arxiv.org/abs/1706.06083}
\BIBentrySTDinterwordspacing

\bibitem{41}
\BIBentryALTinterwordspacing
N.~Carlini and D.~Wagner, ``Towards evaluating the robustness of neural networks,'' 2017. [Online]. Available: \url{https://arxiv.org/abs/1608.04644}
\BIBentrySTDinterwordspacing

\bibitem{43}
A.~Agarwal and N.~K. Ratha, ``Black-box adversarial entry in finance through credit card fraud detection.'' in \emph{CIKM Workshops}, 2021.

\bibitem{44}
Y.~Zhou, M.~Kantarcioglu, B.~Thuraisingham, and B.~Xi, ``Adversarial support vector machine learning,'' in \emph{Proceedings of the 18th ACM SIGKDD International Conference on Knowledge Discovery and Data Mining}.\hskip 1em plus 0.5em minus 0.4em\relax ACM, 2012, pp. 1059--1067.

\bibitem{lunghi2024assessing}
D.~Lunghi, A.~Simitsis, and G.~Bontempi, ``Assessing adversarial attacks in real-world fraud detection,'' in \emph{2024 IEEE International Conference on Web Services (ICWS)}.\hskip 1em plus 0.5em minus 0.4em\relax IEEE, 2024, pp. 27--34.

\bibitem{tiwari2021credit}
P.~Tiwari, S.~Mehta, N.~Sakhuja, J.~Kumar, and A.~K. Singh, ``Credit card fraud detection using machine learning: a study,'' \emph{arXiv preprint arXiv:2108.10005}, 2021.

\bibitem{LRccfd1}
T.~Wang and Y.~Zhao, ``Credit card fraud detection using logistic regression,'' in \emph{2022 International Conference on Big Data, Information and Computer Network (BDICN)}, 2022, pp. 301--305.

\bibitem{LRccfd2}
A.~Mahajan, V.~S. Baghel, and R.~Jayaraman, ``Credit card fraud detection using logistic regression with imbalanced dataset,'' in \emph{2023 10th international conference on computing for sustainable global development (iNDIACom)}.\hskip 1em plus 0.5em minus 0.4em\relax IEEE, 2023, pp. 339--342.

\bibitem{LRccfd3}
M.~V. Krishna and J.~Praveenchandar, ``Comparative analysis of credit card fraud detection using logistic regression with random forest towards an increase in accuracy of prediction,'' in \emph{2022 International Conference on Edge Computing and Applications (ICECAA)}.\hskip 1em plus 0.5em minus 0.4em\relax IEEE, 2022, pp. 1097--1101.

\bibitem{23}
M.~P. LaValley, ``Logistic regression,'' \emph{Circulation}, vol. 117, no.~18, pp. 2395--2399, 2008.

\bibitem{26}
L.~Breiman, ``Random forests,'' \emph{Machine learning}, vol.~45, pp. 5--32, 2001.

\bibitem{27}
S.~J. Rigatti, ``Random forest,'' \emph{Journal of Insurance Medicine}, vol.~47, pp. 31--39, 2017.

\bibitem{45}
\BIBentryALTinterwordspacing
{M. L. G. ULB}, ``Credit card fraud detection,'' n.d. [Online]. Available: \url{https://www.kaggle.com/datasets/mlg-ulb/creditcardfraud/data}
\BIBentrySTDinterwordspacing

\bibitem{46}
\BIBentryALTinterwordspacing
N.~V. Chawla, K.~W. Bowyer, L.~O. Hall, and W.~P. Kegelmeyer, ``Smote: Synthetic minority over-sampling technique,'' \emph{Journal of Artificial Intelligence Research}, vol.~16, p. 321–357, Jun. 2002. [Online]. Available: \url{http://dx.doi.org/10.1613/jair.953}
\BIBentrySTDinterwordspacing

\bibitem{scikit-learn}
\BIBentryALTinterwordspacing
F.~Pedregosa, G.~Varoquaux, A.~Gramfort, V.~Michel, B.~Thirion, O.~Grisel, M.~Blondel, P.~Prettenhofer, R.~Weiss, V.~Dubourg, J.~Vanderplas, A.~Passos, D.~Cournapeau, M.~Brucher, M.~Perrot, and E.~Duchesnay, ``Scikit-learn: Machine learning in python,'' \emph{Journal of Machine Learning Research}, vol.~12, pp. 2825--2830, 2011. [Online]. Available: \url{https://jmlr.org/papers/v12/pedregosa11a.html}
\BIBentrySTDinterwordspacing

\bibitem{art}
\BIBentryALTinterwordspacing
M.-I. Nicolae, M.~Sinn, M.~N. Tran, B.~Buesser, A.~Rawat, M.~Wistuba, V.~Zantedeschi, N.~Baracaldo, B.~Chen, H.~Ludwig, I.~M. Molloy, and B.~Edwards, ``Adversarial robustness toolbox v1.0.0,'' 2019. [Online]. Available: \url{https://arxiv.org/abs/1807.01069}
\BIBentrySTDinterwordspacing

\bibitem{49}
F.~Tramèr, A.~Kurakin, N.~Papernot, I.~Goodfellow, D.~Boneh, and P.~McDaniel, ``Ensemble adversarial training: Attacks and defenses,'' \emph{arXiv preprint}, vol. arXiv:1705.07204, 2017.

\bibitem{50}
G.~K. Dziugaite, Z.~Ghahramani, and D.~M. Roy, ``A study of the effect of jpg compression on adversarial images,'' \emph{arXiv preprint}, vol. arXiv:1608.00853, 2016.

\end{thebibliography}
\end{document}